\def\ii{\'{\i }}

\input{aipcheck}

\documentclass[
    ,final                   
  ]
  {aipproc}

\usepackage{amsmath}

\layoutstyle{8x11double}

\begin{document}

\title{Rapidity long range correlations, parton percolation and color
glass condensate.}

\classification{\texttt{21.65.Qr}}

\author{I. Bautista$^{\dag}$, J. Dias de Deus}{
  address={CENTRA,
Departamento de
F\ii sica, IST, Av. Rovisco Pais, 1049-001 Lisboa, Portugal}
}\author{C. Pajares}{
  address={IGFAE and Departamento
de F\'isica de Part\'iculas, Univ. of Santiago de Compostela,
15782, Santiago de Compostela, Spain} 
}

\begin{abstract}
The similarities between string percolation and Glasma results are 
emphasized, special attention being paid to rapidity long range 
correlations, ridge structure and elliptic flow. As the string density of 
high multiplicity $pp$ collisions at LHC energies has similar value as the 
corresponding to $Au-Au$ semi-central collisions at RHIC we also expect in 
pp collisions long rapidity correlations and ridge structure, extended 
more than 8 units 
in rapidity.
\end{abstract}

\maketitle

\section{}

The mechanisms of parton saturation, string fusion and percolation
have been quite successful in describing the basic facts, obtained
mostly at RHIC, of the physics of QCD matter at high density. Here
we would like to emphasize the similarities between both
approaches string percolation [1][2] and color glass condensate
(CGC) or glasma [3][4] and to show some predictions on rapidity
long range correlations ridge structures [5][6][7][8], width of
the normalized multiplicity distributions and elliptic flow [9].

In string percolation the relevant parameter is the transverse
string density $\eta$, $\eta \equiv N_{s}\pi r_{0}^{2}/\pi R^{2}$ where
$N_{s}$ is the number of strings,  $\pi r_{0}^{2}$ the radial size
of a single string and $\pi R^{2}$ the transverse size of the
collision. $\eta$ behaves like $N_{part}^{2/3}$ and $s^{\Delta}$.
String percolation leads to reduction of particle density at
mid-rapidity, and because of energy-momentum conservation, to an
increase of the rapidity length of the effective strings. The
basic formulae are, for particle density and for $<p_{T}^{2}>$
\begin{equation}
\frac{dn}{dy}=F(\eta)N^{s}\mu \mbox{ and  }
<p_{T}^{2}>=<p_{T}^{2}>_{1}/F(\eta)
\end{equation}
where $F(\eta)$ is the color reduction factor, $F(\eta) \equiv
\sqrt{(1-e^{-\eta})/\eta}$, and $\mu$ and $<p_{T}^{2}>_{1}$ are
the particle density and the averaged transverse momentum squared
of the single string $(<p_{T}^{2}>_{1}r_{0}^{2}\simeq
\frac{1}{4})$, $\frac{1}{F(\eta)}$ or more specifically in the large
$\eta$ limit, $\sqrt{\eta}$, plays the role of the saturation
scale of CGC. Indeed, from (1), the transverse size correlation is
$r_{0}^{2}F(\eta)$ and $1/Q_{s}^{2}$ the corresponding one in CGC.

As far as color electric field is concerned the effective strings
can be identified with the flux tubes of the Glasma picture. In
fact the area occupied by the strings divided by the area of the
effective string gives the number of effective strings [5][6],
\begin{equation}
<N>=\frac{(1-e^{-\eta})R^{2}}{F(\eta)r_{0}^{2}}=(1-e^{-\eta})^{1/2}\sqrt{\eta}(R/r_{0})^{2}
\end{equation}
which in the high density limit is equivalent to $Q_{s}^{2}R^{2}$ the 
number
of flux tubes of the Glasma.

In the process of fusion of strings one has to take care of the
energy momentum conservation, which implies an increase in the
length in rapidity of the string, thus the length of a cluster of
$N_{s}$ strings is given by
\begin{equation}
\Delta y_{N^{s}}=\Delta y_{1}+2 ln N^{s}
\end{equation}
One further notes that overall conservation of energy$/$ momentum regimes for
the number of strings to behave as
\begin{equation}
N^{s}\simeq s^{\lambda}\simeq e^{2 \lambda Y}
\end{equation}
where $Y$ is the beam rapidity, $Y=ln(\sqrt{s}/m)$ and
$\lambda=2/7$. Therefore
\begin{equation}
\Delta y_{Ns}\simeq 2\lambda \Delta Y
\end{equation}
As in the CGC the length in rapidity of the classical fields is
$1/\alpha_{s}(Q_{s}^{2})$ and as the saturation scale is power
behaved in $Y$, we end up in a formula of the kind of (5).
Therefore longitudinal extension of the cluster of strings in
string percolation as in the CGC, increases with energy as $log (s)$,
giving rise to long range rapidity correlations, even for pp at
LHC. Notice that the transverse string density for $pp$ and $AA$
collisions are related by $\eta_{AA}=\eta_{pp}N_{A}^{2/3}$.
Comparing $AA$ collisions at RHIC with $pp$ collisions at LHC, the
factor $N_{A}^{2/3}$ can be balanced with the larger energy of LHC
and the selection of high multiplicity $pp$ collisions. In fig 1.
we show our predictions for the parameter $b$,
$b=(<n_{F}n_{B}>-<n_{F}><n_{B}>)/(<n_{F}^{2}>-<n_{F}>^{2})$ as a
function of the rapidity gap $\Delta \eta$, for $pp$ at LHC
energies, and for a forward on rapidity bins of 0.2. We observe
large rapidity correlations, covering 8 units of rapidity. this
long range rapidity correlations has to do with the ridge
structure observed at RHIC for $A-A$ collisions [5][10]. In fact, the
quantity measured, $\Delta \rho / \sqrt{\rho_{ref}}$ is the
density of particles correlated with a particle emitted at zero
rapidity. The quantity $\Delta \rho$ is the difference in
densities between single events and mixed events, $\rho_{ref}$
coming from mixed samples. It has been shown that [7]
\begin{equation}
\frac{\Delta \rho}{\sqrt{\rho_{ref}}}=R\frac{dn}{dy}F(\phi)
\end{equation}
where $F(\phi)$ describes the azimuthal dependence taken from an
independent model, and $R$ is the normalized 2- particle
correlation given by
\begin{equation}
\begin{split}
R\equiv\frac{<n^{2}>-<n>^{2}-<n>}{<n>^{2}} \\
=\frac{<N^{2}>-<N>^{2}}{<N>^{2}}=\frac{1}{k}
\end{split}
\end{equation}
being $1/k$ the normalized fluctuations of the number of effective
strings (color flux tube in the Glasma) distribution.
If the particle distribution is negative binomial with a $NB$ parameter 
$k_{NB}$ as in string percolation or in glasma, then $k\equiv k_{NB}$.

In the low density regime the particle density is essentially Poisson and 
we have 
\begin{equation}
\eta \rightarrow 0 \mbox{   ,   } k\rightarrow \infty
\end{equation}

In the large $\eta$, large $<N>$ limit, if one assumes that the 
N-effective behave like a single string, with $<N^{2}>-<N>^{2} \simeq 
<N>$, one obtains
\begin{equation}
\eta \rightarrow \infty \mbox{  ,  } k \rightarrow <N> \rightarrow \infty
\end{equation}

An important consequence of (7) and (9), if one assumes that the effective 
strings in the high energy limit emit particles as independent sources, 
is that $k_{1}$ for the single effective string is given by
\begin{equation}
k_{1} \equiv \frac{k}{<N>} \simeq 1
\end{equation}
corresponding to Bose-Einstein distribution. That behaviour was predicted 
out before in the Glasma [6], as an amplification of the intensity of 
multiple emitted gluons.

A parametrization fo $k$ satisfying  (8) and (9) is 
\begin{equation}
k \simeq \frac{<N>}{(1-e^{-\eta})^{3/2}}
\end{equation}
(In the CGC, $k=<N>$, being implicit that the relation only works for the 
high density regime). Then
\begin{equation}
R \frac{dn}{dy}=\frac{1}{k} \frac{dn}{dy}=(1-e^{(-\eta)})^{3/2} \mu
\end{equation}
On the other hand as
\begin{equation}
\frac{1}{k}\frac{dn}{dy} = \frac{1}{\frac{1}{b}-1}
\end{equation}
the range in rapidity of the correlation observed in $b$, is the same of 
the rapidity range of the ridge structure. In this way, it was predicted 
at LHC a ridge structure for high multiplicity $pp$ collisions [5][9][10], 
extended 8 units of rapidity. The CMS collaboration has reported recently 
such ridge structure [11].

From (7), $1/k$ is the width of the distribution $<n>P_{n}$ 
as a function of $n/<n>$. The curve $k$ as a function of $\eta$ shows a 
minimum at $\eta \simeq 1.2$. At low density we have $k$ decreasing with 
$\eta$ and a larger density we have $k$ increasing with $\eta$. This 
change of behavior of $k$ can be reached for pp at the highest 
energy of 
LHC and therefore the distribution $<n>P_{n}$ will start to be narrower 
[10].

Finally, let us mention that string percolation describes rightly the 
observed elliptic flow, and its dependence on $p_{T}$, rapidity and 
centrality [9]. In percolation are obtained close analytical expressions 
for a 
$v_{2}(p_{T})$. For instance for the $p_{T}$ integrated elliptic flow
$v_{2}$ we obtain [9]
\begin{equation}
v_{2} \equiv \frac{e^{-\eta}-F(\eta)^{2}}{2F(\eta)^{3}}\frac{R}{R-1} 
\frac{2}{\pi}\int^{\pi/2}_{0} cos(2 \varphi)(\frac{R_{\varphi}}{R})^{2}
\end{equation}
where
\begin{equation}
R_{\varphi}=R \frac{sin(\varphi -\alpha)}{sin \varphi}
\end{equation}
\begin {equation}
\alpha = sin^{-1}(\beta sin \varphi )
\end{equation}
and $\beta = \frac{b}{2R}$, being $b$ the impact parameter.
In fig 2 we compare our results for $Au-Au$ collisions at $\sqrt{s}=200$ 
GeV, for $N_{part}=211$ with the experimental data for the full range of
pseudorapidity $\eta$.
In fig. 3 we show our prediction for $Pb-Pb$ collision at $\sqrt{s}=5.5$ 
TeV.

Summarizing up, string percolation and glasma, which have been quite
successful in describing the basic experimental facts obtained at RHIC,
give defined predictions for LHC energies on rapidity long range
correlations, ridge structure and on the width of the $<n>P_{n}$
distributions.

Such predictions, which are valid not only for $AA$ collisions but also
for $pp$ high multiplicity collisions, if confirmed by the experimental
data, can help to establish the existence of a highly coherent state
formed of strong color field extended several units of rapidity.

\begin{figure}
  \includegraphics[height=.28\textheight]{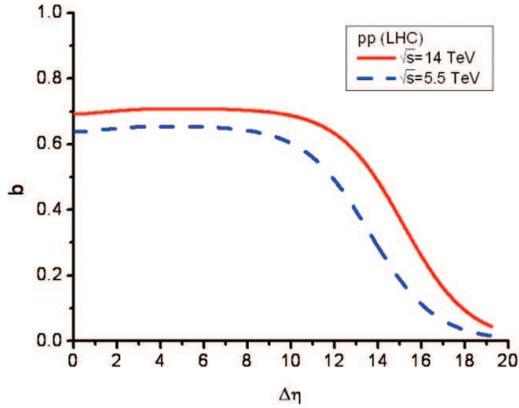}
  \caption{The normalized forward-backward dispersion $b$ as a function of 
the rapidity gap, $\Delta \eta$, between the forward and backward bins, 
for $pp$ collisions. The bins are of 0.2 units of rapidity. Solid line, 
$\sqrt{s}=14$ TeV, dashed line $\sqrt{s}=7$ TeV.}
 \end{figure}

\begin{figure}
  \includegraphics[height=.23\textheight]{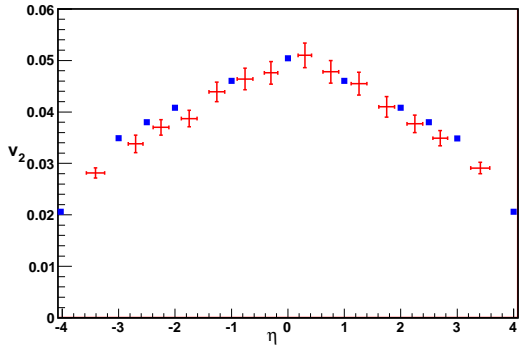}
  \caption{$v_{2}$ as a function of pseudorapidity for $N_{part}=211$ in 
Au-Au collisios at $\sqrt=200$ GeV. Dots are our results, and the data 
is taken from reference [12].
} \end{figure}

\begin{figure}
  \includegraphics[height=.23\textheight]{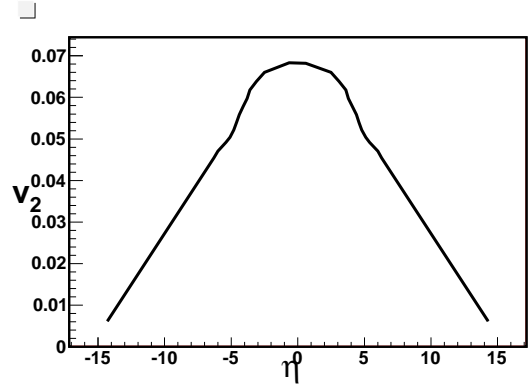}
  \caption{$v_{2}$ as a function of pseudorapidity for central $Pb-Pb$ 
collisions at $\sqrt{s}= 5.5$ TeV.
} \end{figure}

\begin{theacknowledgments}
 C. P. thanks the organizers for the nice meeting. This work is under the 
project FPA2008-01177 of Spain and the Xunta de Galicia. J.D.D. thanks
the support of the FCT/Portugal project PPCDT/FIS/575682004.
\end{theacknowledgments}

\bibliographystyle{aipproc}

\end{document}